\documentclass[pdflatex,sn-nature]{sn-jnl}

\usepackage{graphicx}%
\usepackage{multirow}%
\usepackage{amsmath,amssymb,amsfonts}%
\usepackage{amsthm}%
\usepackage{mathrsfs}%
\usepackage[title]{appendix}%
\usepackage{xcolor}%
\usepackage{textcomp}%
\usepackage{manyfoot}%
\usepackage{booktabs}%
\usepackage{algorithm}%
\usepackage{algorithmicx}%
\usepackage{algpseudocode}%
\usepackage{listings}%

\begin{document}

\title{Transfer printing micro-assembly of silicon photonic crystal cavity arrays: beating the fabrication tolerance limit}

\author*[1]{\fnm{Sean P.} \sur{Bommer}}\email{sean.bommer@strath.ac.uk}
\author[2]{\fnm{Christopher} \sur{Panuski}}
\author[1]{\fnm{Benoit} \sur{Guilhabert}}
\author[1]{\fnm{Zhongyi} \sur{Xia}}
\author[1]{\fnm{Jack A.} \sur{Smith}}
\author[1]{\fnm{Martin D.} \sur{Dawson}}
\author[2]{\fnm{Dirk} \sur{Englund}}
\author*[1]{\fnm{Michael J.} \sur{Strain}}\email{michael.strain@strath.ac.uk}

\affil[1]{Institute of Photonics, Dept. of Physics, University of Strathclyde, UK}
\affil[2]{Research Laboratory of Electronics, MIT, USA}

\abstract{Photonic crystal cavities (PhCCs) can confine optical fields in ultra-small volumes, enabling efficient light-matter interactions for quantum and non-linear optics, sensing and all-optical signal processing.  The inherent nanometric tolerances of micro-fabrication platforms can induce cavity resonant wavelength shifts two-orders of magnitude larger than cavity linewidths, prohibiting fabrication of arrays of nominally identical devices.  We address this device variability by fabricating PhCCs as releasable pixels that can be transferred from their native substrate to a receiver where ordered micro-assembly can overcome the inherent fabrication variance.  We demonstrate the measurement, binning and transfer of 119 PhCCs in a single session, producing spatially ordered arrays of PhCCs, sorted by resonant wavelength.  Furthermore, the rapid in-situ measurement of the devices enables measurements of the PhCCs dynamic response to the print process for the first time, showing plastic and elastic effects in the seconds to hours range.}

\maketitle

\section{Introduction}
Two dimensional photonic crystal cavities allow the extreme confinement of light within solid state materials.  Their ultra-high spatial confinement, together with resonant quality factors (Q-factors) that can exceed $10^6$ in common materials such as silicon \cite{Noda_PHCC_2018,Galli_Mem_PhCC_2014}, allow for significant enhancement of light-matter effects necessary for sensing\cite{Deasy2014,Picelli2020}, non-linear optics\cite{Rossi_OPO_2021,chopin_ultra-efficient_2023}, ultra-compact lasers\cite{Noda_laser_2017,Yvind_laser_2022} and coupling with single photon emitters\cite{Arakawa_qdot_2020,Wilson_QdotCav_2024,Petruzzella2017}.  By using a suspended device geometry\cite{Noda_membrane_2012} the optical mode can be strongly confined to the semiconductor material with modal confinement provided by arrays of etched holes in the plane of the cavity, and high refractive index contrast with air above and below the device.  For devices with resonant wavelengths in the telecommunications spectral band, the etch hole dimensions are on the order of $150$\,nm in diameter, through a suspended membrane thickness of $220$\,nm, ensuring a single supported optical mode in the vertical dimension.  

The ultra-high optical confinement of these devices, and the sub-micron scale of their critical geometric features, makes them highly susceptible to variations in the fabrication process.  Although PhCCs have been demonstrated using deep UV lithography methods\cite{Krauss_DUV_2006,AIM_DUV_2022,Ooka2015}, the highest level of accuracy in defining their physical geometry can be obtained by using electron beam lithography, where typical fabrication variances are in the few nanometre range, after both lithography and etching stages\cite{samarelli2008optical,Thomas_PCstats2011}.  The physical variation in these critical features can in turn lead to variation in the device resonant wavelength that can be in the few nanometre range, within the $1550$\,nm spectral region\cite{Noda_var_2011,Minkov_var_2013}.  For devices with linewidths that can be over 2 orders of magnitude less than this variation, direct fabrication of nominally identical devices becomes impossible.  In many applications, use of a single resonator, or identification of 'hero performance' is sufficient, and so particular devices can be selected from an array where properties will vary.  However, as PhCCs begin to be applied in cases where ensembles\cite{Panuski_SLM2022}, or coupling between co-resonant cavities are required, it is imperative to move beyond the fabrication limitations of the platform.  One recent approach employed laser enhanced selective oxidation of an array of cavities after initial spectral characterisation to align the resonant wavelength of $64$ PhCCs to within a standard deviation of $2.5$\,pm\cite{Panuski_SLM2022}.  This actively monitored trimming process allowed an ensemble of devices to operate on a single optical pump beam for high speed spatial light modulation.  Post-fabrication tuning of individual PhCCs has also been demonstrated using mechanical \cite{midolo_electromechanical_2011,Petruzzella2017}, cladding refractive index \cite{Monat_tuning_2010,Brossard_Ink_2017}, thermal \cite{Chong_thermal_2004} and electronic \cite{husko_ultrafast_2009} tuning mechanisms, though routes to scaling these for dense arrays of cavities are challenging.  

In this work we present a physical transfer method where, rather than tuning cavities in a fixed spatial arrangement, the individual devices are fabricated as mechanically separable pixels which can then be characterised, binned and physically rearranged onto a new substrate.  This represents two major challenges in the creation of arrays fabricated from single pixel devices.  Firstly, the PhCC pixel devices must be detachable from their native substrate and transferred onto a receiver substrate, whilst maintaining a suspended geometry to preserve their optical characteristics.  Secondly, the process of optically characterising the devices needs to be carried out within the same system as the transfer process to enable high-throughput of device binning and assembly.  This latter point is critical as the majority of high-accuracy transfer techniques rely on physically shuttling samples between characterisation and micro-assembly systems.  Here, by integrating both measurement and transfer assembly processes into a single system, we not only enable the deterministic selection and transfer of 119 devices in a single session, but also unlock the possibility of measuring dynamic changes in device performance during the printing process that are inaccessible using traditional serial integration and measurement methods.

We present the measurement and subsequent transfer of 119 silicon PhCC devices into an array ordered by resonant wavelength.  The transfer printing process is shown to be repeatable with single pixel devices being transferred up to 5 times before resonant wavelength shifts beyond their linewidth are induced.  A swept wavelength spectral measurement system is integrated with the transfer printing tool to enable in-situ optical characterisation.  Dynamic effects of the printing process on the cavities' resonant wavelengths are measured on the seconds to hours timescale, uncovering elastic effects in their response.

\section{Transfer printing of PhCCs and in-situ spectral measurement}
\begin{figure}[h]
    \centering
    \includegraphics[width=\textwidth]{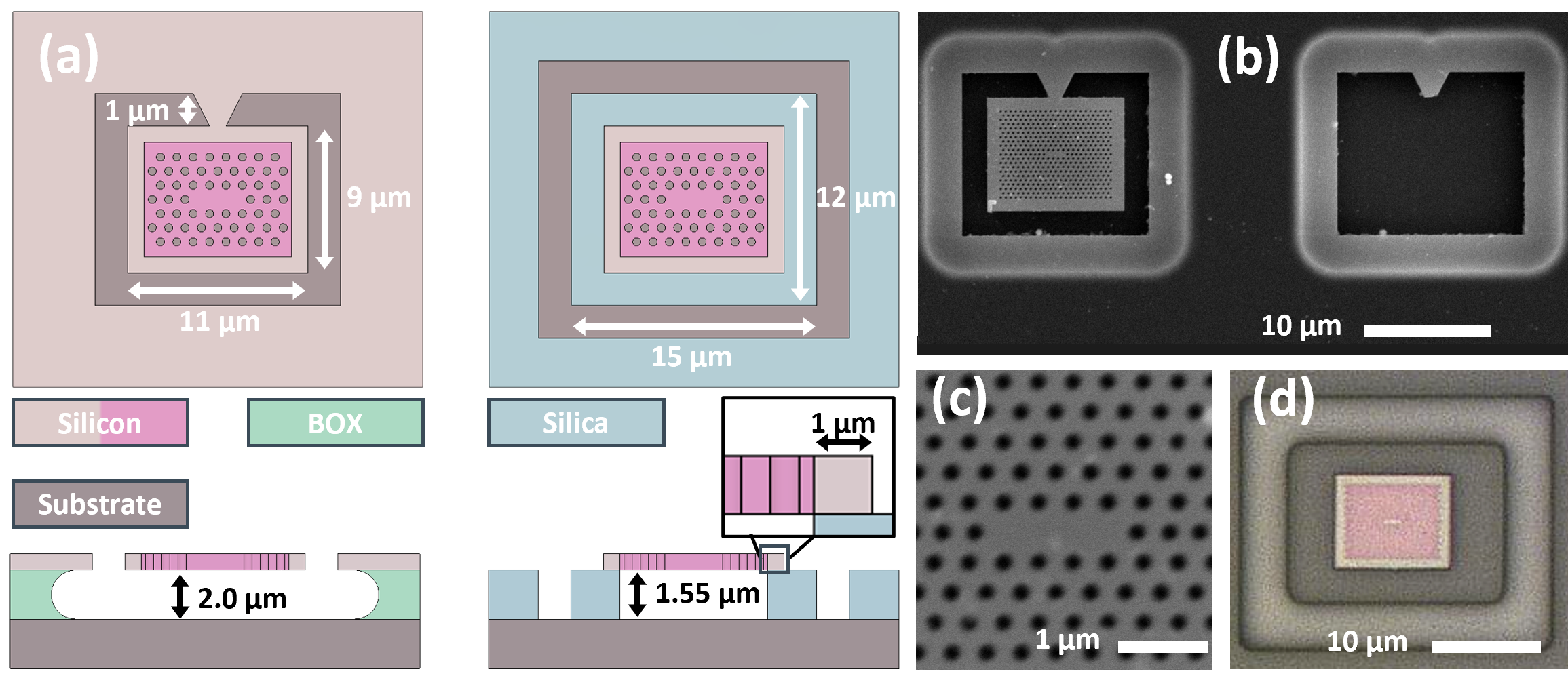}
    \caption{\textbf{Releasable PhCC pixels for mechanical transfer.} (a) Schematic of a PhCC pixel on its donor substrate and after transfer to a receiver substrate with silica support frame. (b) SEM image of fabricated PhCC pixels on the donor substrate, where the right hand pixel has been transferred, leaving a void in the donor array. (c) High magnification SEM image of the PhCC central region showing the L3 cavity geometry and (d) optical microscope image of a silicon PhCC printed onto a silica support frame.}
    \label{fig:CavitySchematic}
\end{figure}

The process for transfer printing of micron-scale membrane photonic devices is well established\cite{smith_hybrid_2022,Carlson2012b,Corbett2017,Katsumi2018}, and typically involves the use of a soft polymer stamp to pick-up, align and place devices onto host substrates.  The vast majority of work has focussed on the printing of membrane devices onto planar substrates where contact is made across the full membrane surface \cite{Cho2016,Loi2018,Karnadi2014,Ghaffari2010,Picelli2020} or where the contact area of printed devices is in excess of any suspended regions or areas with topological distortion \cite{Carlson2012,Osada2018,Hill2018}.  For PhCC devices, the membranes must be printed with a suspended geometry to preserve the air-cladding above and below the PhCC for strong optical confinement, and replication of performance on its native substrate.  \figurename\,\ref{fig:CavitySchematic} shows a schematic and scanning electron microscope images of our single pixel PhCC devices.  The optical cavity needs to be large enough to maintain the optical confinement of the mode, whilst minimising the total area to enable multiplexing of pixels together on a host substrate.  Furthermore, when releasing micron-scale membrane devices from their host substrate, the physical support structure, or tether, has to be carefully designed to allow cleaving and prevent collapse of the membrane onto its substrate.  In this work, the silicon PhCC devices used follow the design of an L3 cavity based on references \cite{minkov_photonic_2017,Panuski_SLM2022}.  The full photonic crystal area is 7 x 9 $\mu $m$^2$.  To minimise the total pixel area, and achieve high density integration, the external planar membrane border around the PhCC was limited to 1 $\mu m$ in width.  \figurename\,\ref{fig:CavitySchematic}(a) shows a schematic of the printed PhCC in its suspended geometry.  In this work, the PhCC devices were printed directly onto the support structures with no additional adhesion layers.  The lateral overlap between the supporting rectangular frame and PhCC border region of $<1 \mu m$, requiring a transfer print positioning accuracy in the 100's of nanometres range.  As we have previously demonstrated, this positional accuracy is achievable using optical microscopy overlay alignment methods \cite{McPhillimy2020,McPhillimy2018}. Furthermore, by transferring the PhCC devices onto a receiver structure, the separation of the membrane bottom surface and underlying substrate is no longer bound by the format of the fabrication wafer.  In suspended silicon PhCC devices, a $220$\,nm silicon core is suspended over a silicon substrate at a distance of $2\,\mu $m, matching the thickness of buried oxide that is chemically etched, after lithography and reactive ion etching of the PhCC holes structures, to create an under-cladding of air.  \figurename\,\ref{fig:CavitySchematic}(b) shows a scanning electron miroscopy (SEM) imaging of a representative PhCC device suspended from the native silicon substrate (left) beside a void where a PhCC has been transferred (right). \figurename\,\ref{fig:CavitySchematic}(c) is an SEM image close up of the representative L3 cavity of a suspended PhCC on the native silicon substrate.  In the receivers fabricated for this work, we used Plasma Enhanced Chemical Vapour Deposition (PECVD) silica layers of $1.55\,\mu $m thickness on a silicon substrate.  The silica suspension frames were fabricated using optical lithography and reactive ion etching.  The $1.55\,\mu $m separation was deliberately targeted to promote constructive interference between the internal PhCC mode and the reflected light from the air-substrate interface, to improve vertical out-coupling of the cavity towards the microscope measurement optics\cite{kim_vertical-cavities_2012}. An example of a printed PhCC on a receiver silica frame in suspended geometry is shown in \figurename\,\ref{fig:CavitySchematic}(d).

\begin{figure}[h]
    \centering
    \includegraphics[width=\textwidth]{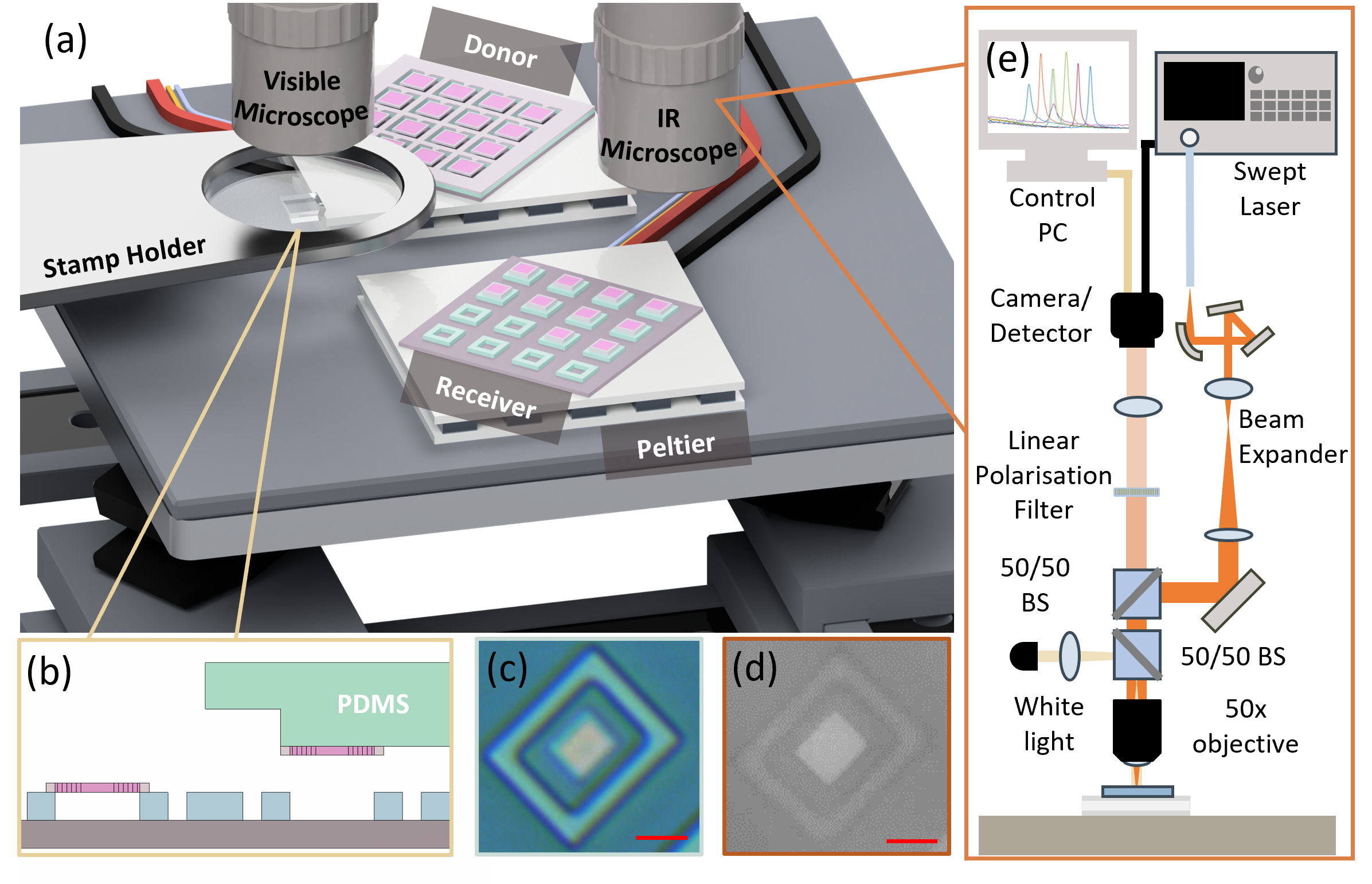}
    \caption{\textbf{Accurate transfer printing system with in-situ reflectivity spectrum measurement capability.} (a) Schematic TP system and optical measurement rig incorporating high-accuracy 6-axis stage, fixed stamp holder, Peltier mounted donor and receiver samples and optical microscope objective lenses.  (b) Schematic of the pixel printing process using a PDMS stamp. (c) and (d) show images of a printed PhCC pixel in the visible and IR systems respectively. (e) Schematic detail of the optical injection and measurement system embedded in the TP tool.}
    \label{fig:TPsystem}
\end{figure}

A schematic of the transfer print system with optical measurement module is presented in \figurename\,\ref{fig:TPsystem}.  The transfer printing system comprises an optical microscope column, fixed transfer stamp holder and a high-accuracy 6-axis translation stage with Peltier coolers, on which the donor and receiver samples are mounted \cite{guilhabert_advanced_2023}.  The Peltier elements are held a few degrees below the cleanroom laboratory temperature at a constant $19.6^o$C with closed loop feedback to ensure stable operation of the PhCCs. In addition to the wide-field optical imaging function of the microscope column, the system also incorporates an optical injection line for the swept-wavelength tuneable laser characterisation system.  The signal reflected from a targeted PhCC is coupled back through the microscope objective to a camera system for assessment of the spatial mode and measurement of the reflectivity spectrum under swept wavelength operation.

\begin{figure}[h]
    \centering
    \includegraphics[width=0.7\textwidth]{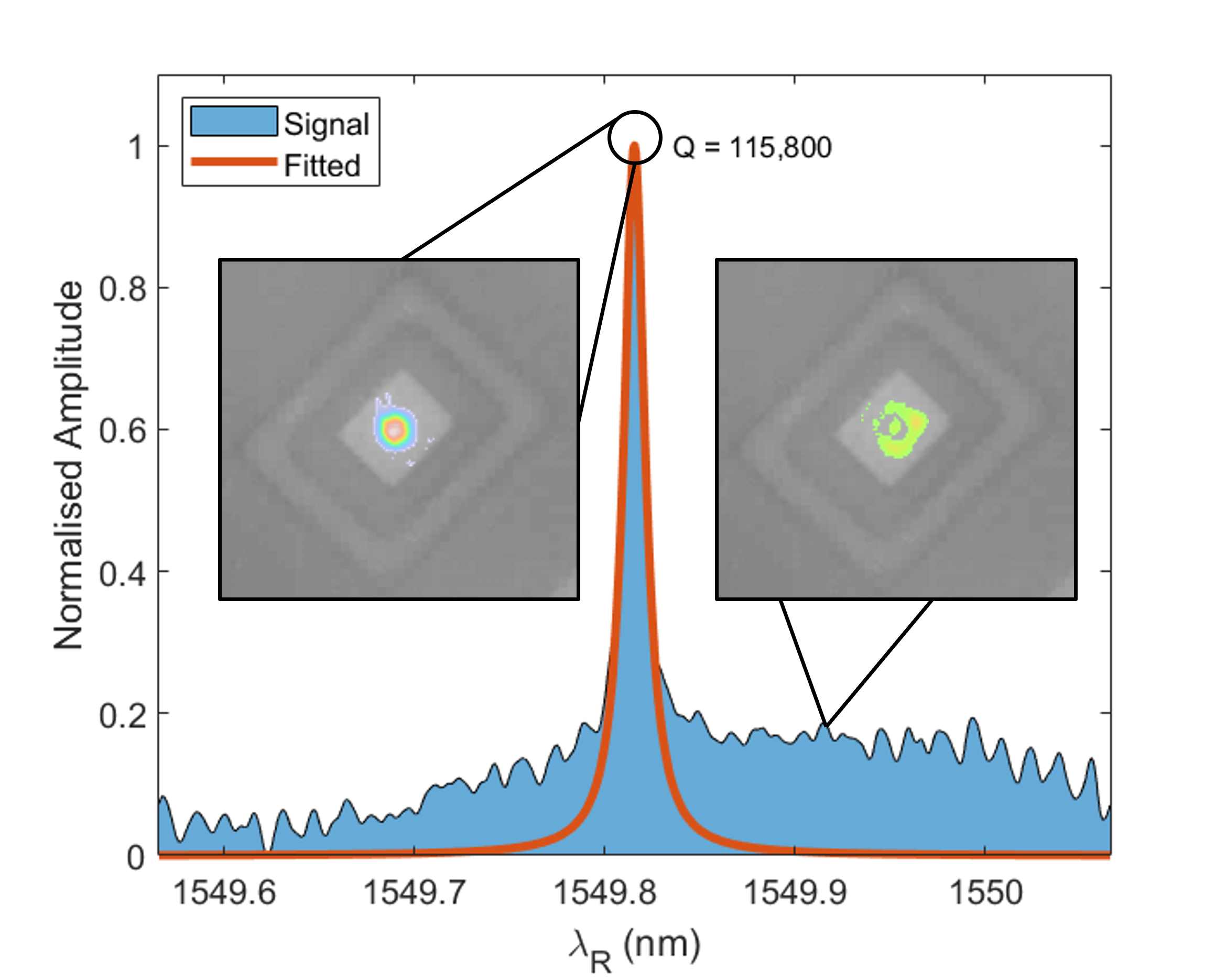}
    \caption{\textbf{Spectral measurement of PhCC in the transfer system.} Measured reflectivity spectrum from a PhCC captured using the in-situ measurement system.  The PhCC is printed on a receiver substrate silica suspension frame. Insets show the spatial mode images captured by the InGaAs camera at two points in the tuneable wavelength sweep corresponding to on-resonance (left) and off-resonance (right) conditions.}
    \label{fig:CavitySpectrum}
\end{figure}

\figurename\,\ref{fig:CavitySpectrum} shows a reflectivity spectrum measurement from a representative PhCC device captured using the in-situ scanning laser measurement system in the transfer print tool.  The central resonant wavelength, linewidth and Q-factor of the resonator can be extracted from this swept laser measurement in real-time, with single pass sweeps taking only a few seconds to complete. Details of the method for parameter extraction from the spectral measurements is presented in appendix A.  For the PhCC's in this work, the resonant wavelengths were in a range around $\sim 1551$\,nm, with cavity linewidths of $\sim 15$\,pm and Q-factors in the $10^5$ range.

\section{Spatial ordering of PhCC array by resonant wavelength}
\begin{figure}[h]
    \centering
    \includegraphics[width=0.95\textwidth]{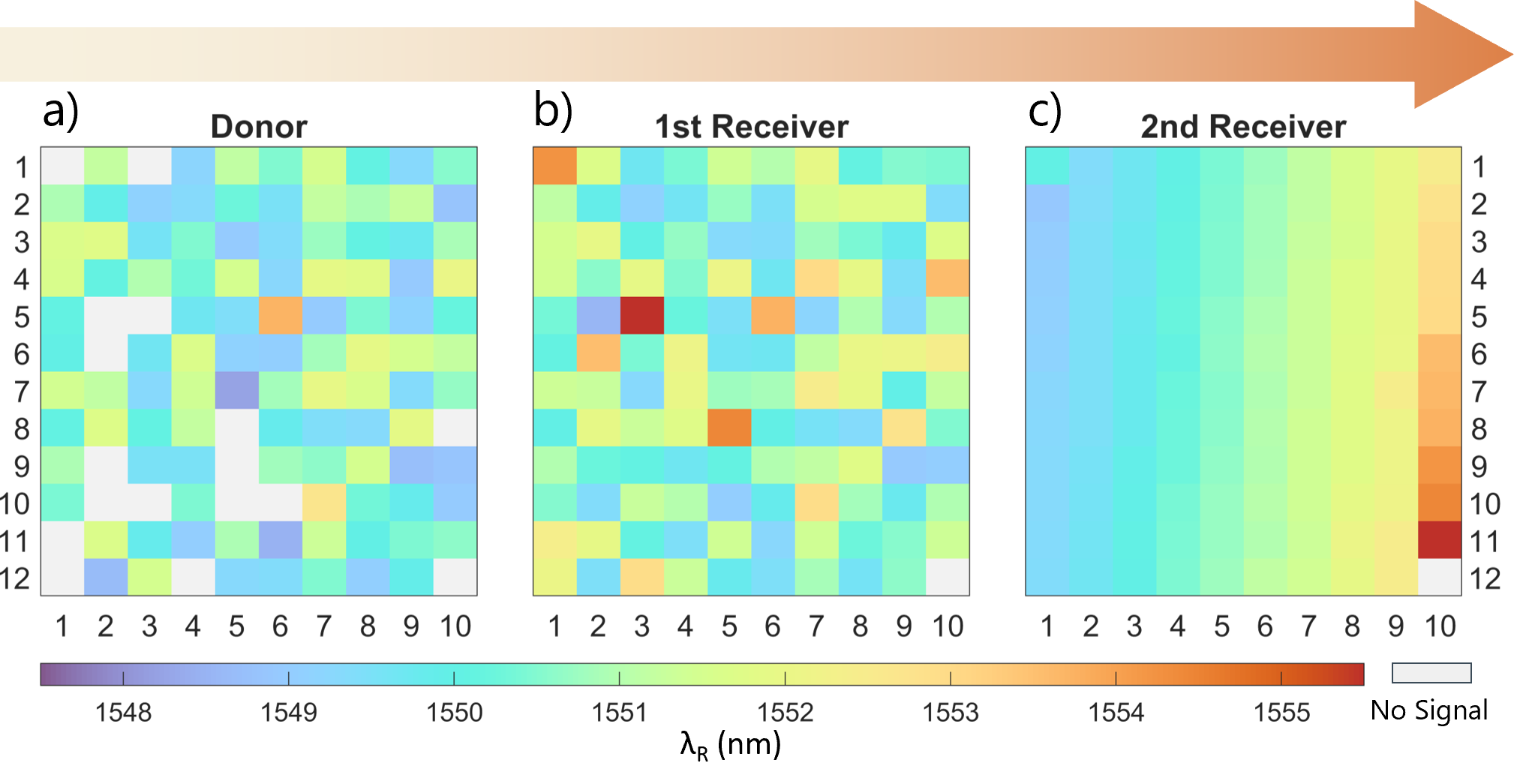}
    \caption{\textbf{Spatial ordering of PhCC arrays by resonant wavelength.}  Measured resonant wavelengths of the PhCCs on (a) donor substrate, (b) 1st receiver and (c) 2nd receiver substrate.}
    \label{fig:WavelengthMaps}
\end{figure}

To enable spatial ordering of an as-fabricated PhCC array, 120 devices were first measured on their native substrate, and the resultant map of PhCC pixel resonant wavelengths is shown in \figurename\,\ref{fig:WavelengthMaps}(a).  As expected, the measured cavity wavelengths span across a range of a few nanometres in wavelength, corresponding to the variations in cavity geometry induced by nano-fabrication geometry tolerances. The blank spots in the figure correspond to PhCCs where no spectral measurements could be obtained, likely due to partial collapse of the pixels onto the exposed substrate during the under-etch process.  One pixel could not be released from the donor chip, so all further results correspond to the remaining 119 pixels that could be transferred. The resonant wavelengths are distributed across the array without any clear spatial pattern, as shown in \figurename\,\ref{fig:WavelengthMaps}(a).  The devices were then numerically sorted by resonant wavelength and printed onto receiver substrate following the spectral ordering.

\begin{figure}[h]
    \centering
    \includegraphics[width=0.8\textwidth]{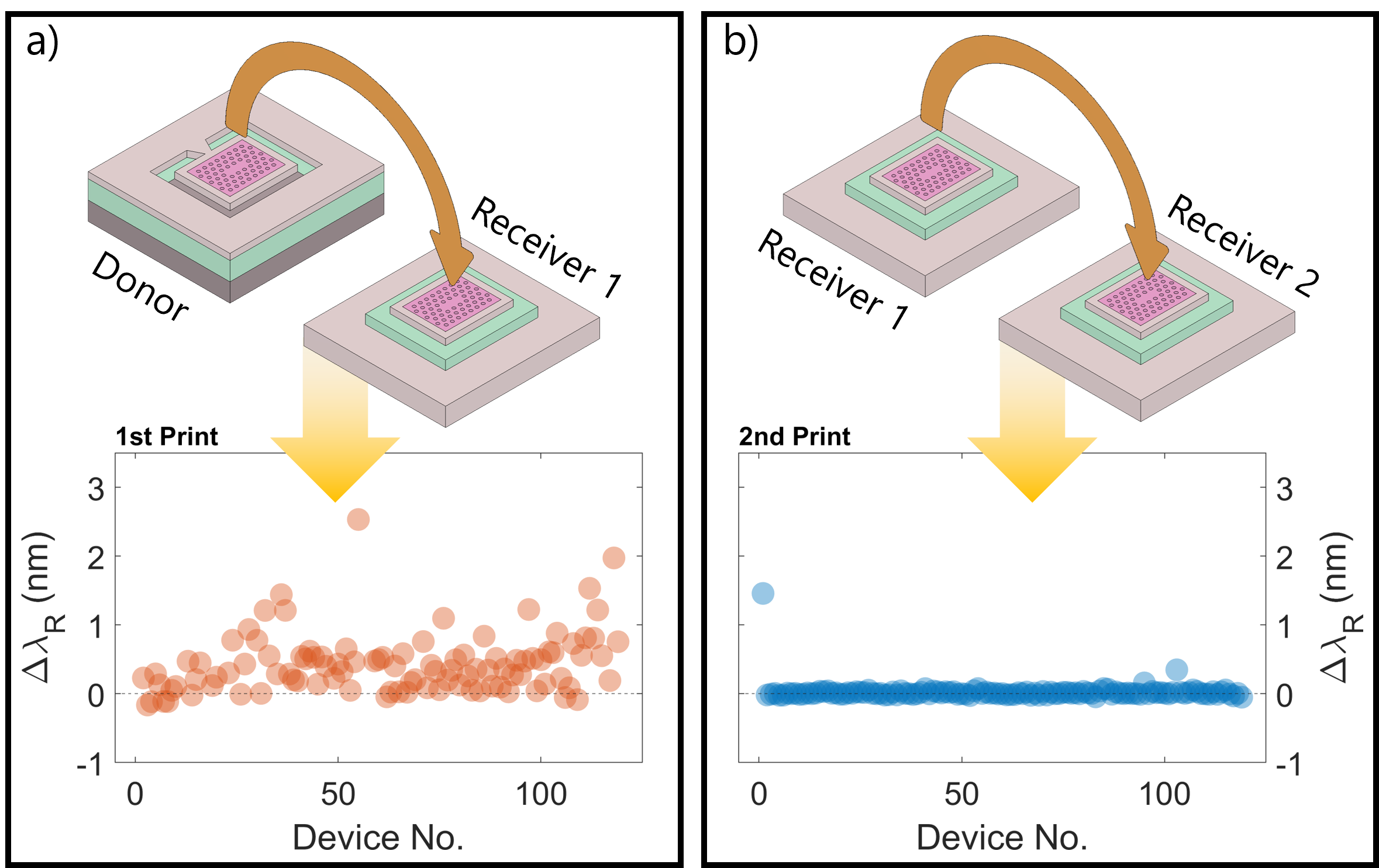}
    \caption{\textbf{Printing induced resonant wavelength shift.} Measured resonant wavelength shifts of individual PhCCs after (a) 1st print, and (b) 2nd print}
    \label{fig:WavelengthScatter}
\end{figure}

\figurename\,\ref{fig:WavelengthMaps}(b) shows a map of the printed pixels' resonant wavelengths on a first receiver substrate.  From the first print it is obvious that the spectral ordering has been frustrated and again the cavities form a disordered array, although all transferred pixels are now visible, recovering the previously unmeasureable cases from the donor substrate.  The measurements on receiver 1 were then used to create an ordered array, through selective placement during an additional transfer step onto a second receiver, the results of which are presented in \figurename\,\ref{fig:WavelengthMaps}(c).  The wavelength ordering was preserved faithfully in this transfer, indicating that the process of cleaving the silicon tethers during the first print induced a plastic shift in the cavity resonant wavelengths.  \figurename\,\ref{fig:WavelengthScatter}(a) and (b) show the wavelength shifts of individual PhCCs for the first and second prints, with a clear improvement in the second print case.   For the first print, the mean wavelength shift of $\pm 0.426$\,nm and standard deviation across the array of $\pm 0.438$\,nm, clearly shows the effect of the release from the donor substrate on the cavities where the spectral ordering has been frustrated by the initial release process.  The second print however shows a mean wavelength shift of $\pm 0.025$\,nm and standard deviation of $\pm 0.139$\,nm, demonstrating stable performance of the PhCCs before and after the print comparable with what can be achieved with post-fabrication trimming. The two outlier values of wavelength shift in the second print array correspond to two cases where mechanical effects during printing caused larger deviations of the resonant wavelength.  Device 1 was printed as a calibration step for the force being applied during the print release stage and a higher than necessary value appeared to cause a residual plastic shift in resonant wavelength.  The second outlier device  was accidentally printed onto a rigid area of the sample before being moved to its target receiver location, again exhibiting a larger deviation of the resonant wavelength. Removing these outliers in the second print gives mean and standard deviation wavelength shifts of $\pm 0.007$\,nm and $\pm 0.021$\,nm respectively. 
By spatially ordering the devices, local clusters of PhCCs with resonant wavelength spread within a cavity linewidth can be achieved, without post-fabrication trimming.  Examples of cavity clusters are presented in appendix B. 

\begin{figure}[h]
    \centering
    \includegraphics[width=0.85\textwidth]{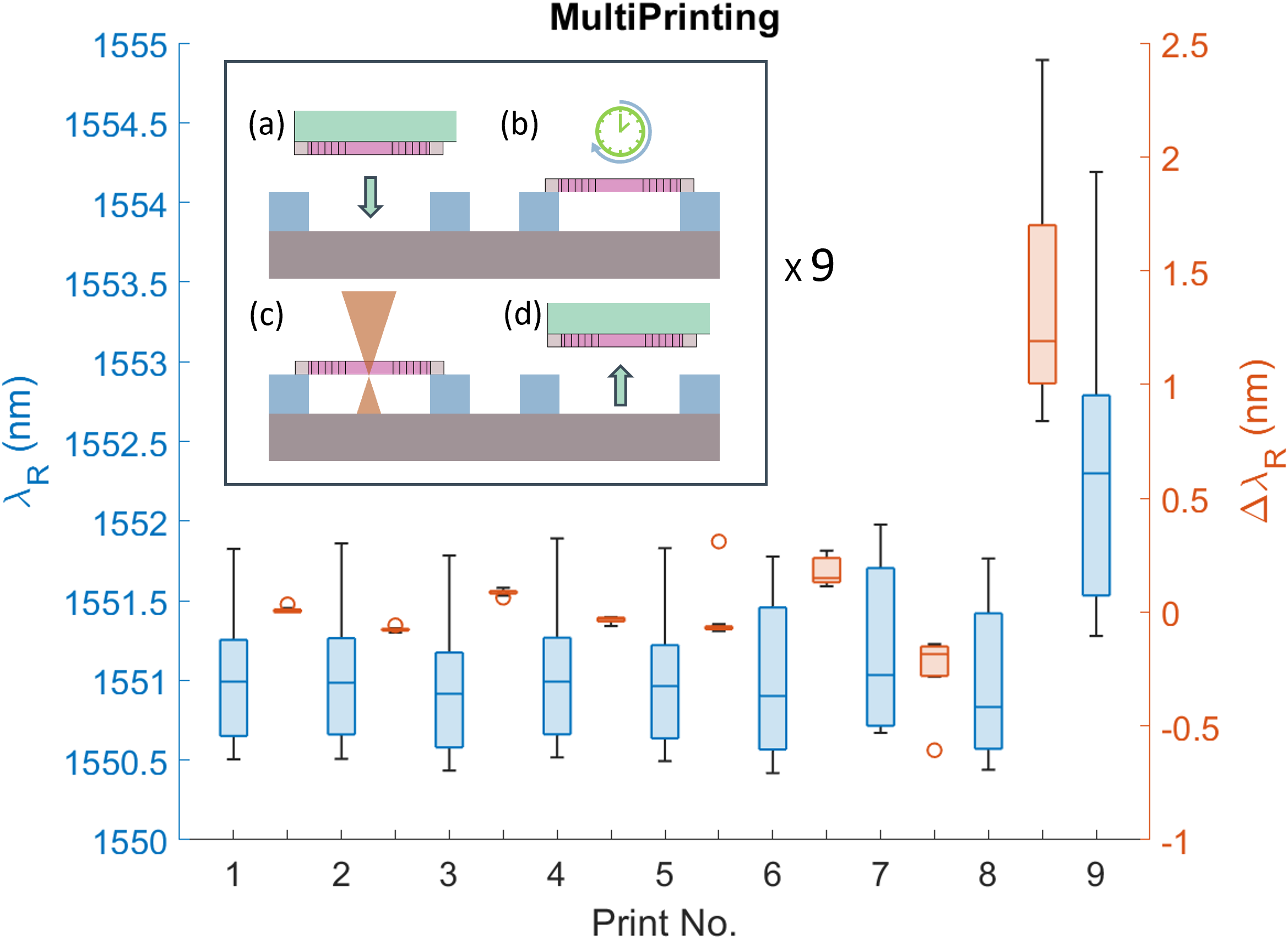}
    \caption{\textbf{Repeated print cycles.} Measured cavity wavelength sets after each printing cycle for a set of 10 PhCC devices, showing absolute cavity resonant wavelength (blue) and relative resonance shift between each print location (red).  (Inset illustrates the print and measurement cycle.)}
    \label{fig:Cavity Multiprinting}
\end{figure}

An additional set of 10 PhCCs from the donor chip, with resonant wavelength around $1550$\,nm were selected to test the effects of multiple print cycles on the cavity response.  \figurename\,\ref{fig:Cavity Multiprinting} shows the measured resonant wavelengths of the PhCCs after each print cycle and the relative shift of individual devices. All print cycles are numbered after the initial print '0' from the donor to receiver to avoid inclusion of the plastic shift effects in these results.  For 5 print cycles, after the initial print from the donor substrate, the subset of PhCCs show negligible wavelength shifts.  The stability then degrades over the next few cycles, reverting to deviations close to the as-fabricated array.  This could be due to the effects of surface contamination from the environment or from the transfer stamp itself, but no clear effects were visible under electron microscope inspection.

\section{Dynamic cavity response measured in-situ}
\begin{figure}[h]
    \centering
    \includegraphics[width=0.65\textwidth]{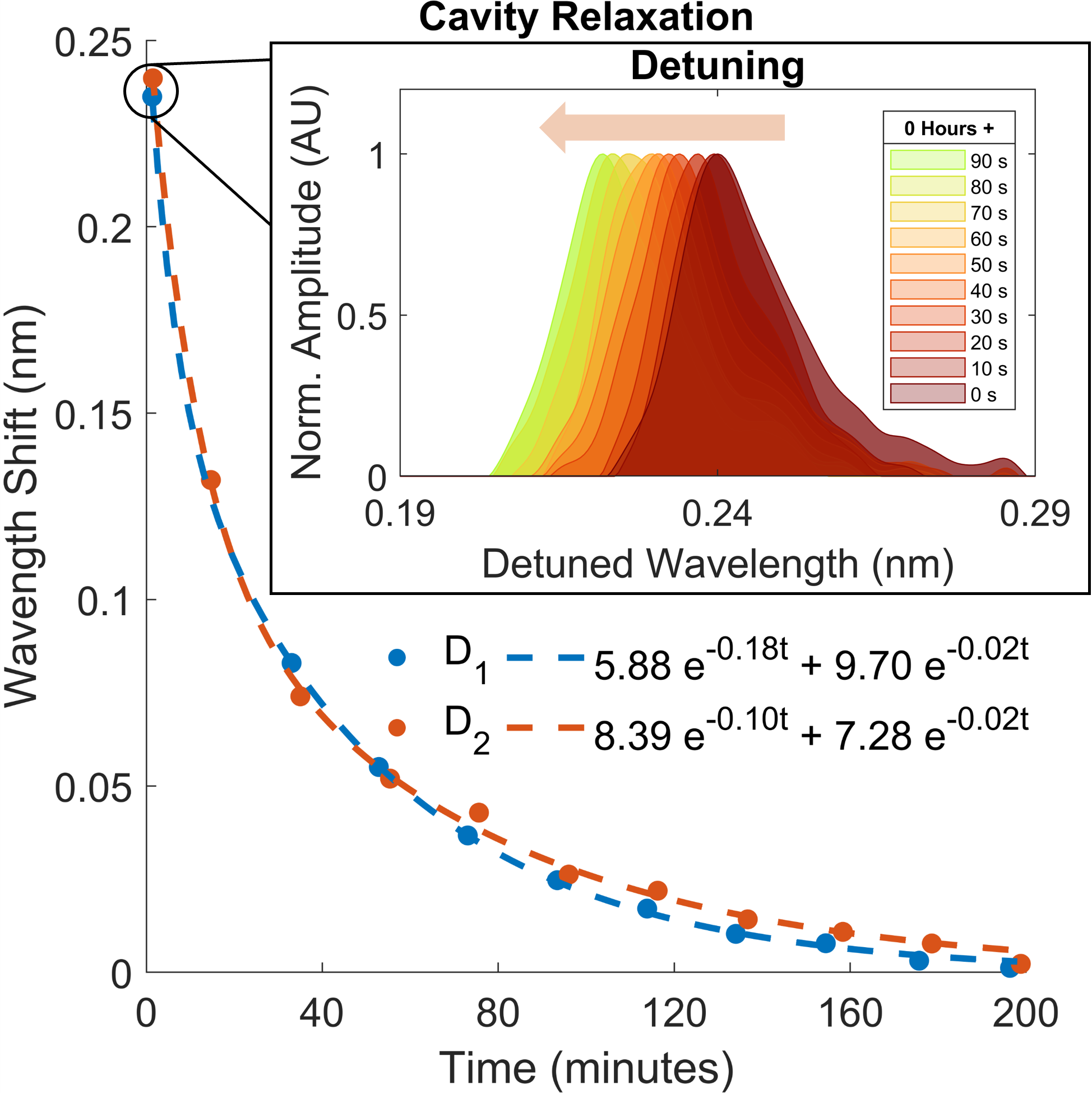}
    \caption{\textbf{Dynamic effects of printing on the cavity optical response.} Cavity resonance wavelength for 2 separate PhCCs measured over a 200 minute period after initial printing.  The inset shows the fast relaxation of the cavity resonance just after printing, as captured by the in-situ measurement and the stable steady state response after relaxation.}
    \label{fig: Cavity Relaxation}
\end{figure}

As detailed in the previous section, the release of the silicon PhCC pixels from their native substrate results in a permanent shift in their cavity resonant wavelength.  Since the spectral measurements of the PhCCs are carried out in-situ in the transfer printing system, dynamic effects can be monitored, limited by the timescale of the swept laser spectral measurement system.  The inset to \figurename\,\ref{fig: Cavity Relaxation} shows the time varying cavity reflectivity spectrum for a single PhCC cavity over a few tens of seconds just after the initial printing, showing a clear shift of the spectrum with time. The main decay curve presented in \figurename\,\ref{fig: Cavity Relaxation} shows the results of spectral measurements of 2 different PhCCs on the same substrate measured concurrently over the course of 200 minutes, plotting the measured resonant wavelength shift relative to the pre-printed steady-state value. The printing process shows a clear red-shift in cavity resonant wavelength immediately after release onto the receiver substrate, followed by a relaxation back towards the steady state value on a timescale of around 200 minutes.    

The results of the multiple print cycles show that steady-state cavity resonances are preserved for up to 5 cycles, demonstrating that the print process does not leave measurable material deposits on the cavity, which would induce a red-shift of the resonance wavelength due to increasing local cladding refractive index.  The dynamic relaxation of resonant wavelengths rather suggests a strain based effect induced by the contact printing that then relaxes as the free-standing PhCCs are left on the suspension frames with no restraining force beyond the local forces between their surfaces.  Furthermore, the relaxation curves can be fitted to a double exponential function, with two characteristic relaxation time constants originating from potentially distinct physical processes.  However, simple measurement of the cavity resonant wavelength does not currently allow further insight into the origins of these effects.

\section{Conclusions}
We have demonstrated a post-fabrication integration method that allows the spectral measurement, binning and spatial arrangement of high Q-factor Photonic Crystal Cavity (PhCC) devices into ordered arrays, with precision beyond what is achievable in as-fabricated device arrays.  The transfer printing integration of 119 PhCC pixels demonstrated handling of devices with dimensions in the few micrometre range, onto suspension frames on receiver substrates with physical contact limited to less than $1\,\mu $m in overlap.  The printing process preserved cavity resonant wavelengths within their linewidth range after a first permanent shift induced by the print release from the donor substrate.  Furthermore, the printing process preserved cavity performance over multiple cycles, allowing for reconfiguration or rework of samples using this method.  The in-situ optical measurement of cavities enabled study of the printing dynamics in the seconds to hours timescale, exhibiting elastic relaxation effects in the cavities after printing that would not have been easily measurable in standard integration and measurement system combinations.  The ability to directly measure micron scale device response and subsequently integrate large numbers of devices onto host chips breaks the dependence on fabrication limited integrated optical systems and paves the way for future, high performance optical systems-on-a-chip.

\section*{Acknowledgements}

The authors acknowledge funding from the following sources: Royal Academy of Engineering (Research Chairs and Senior Research Fellowships); Engineering and Physical Sciences Research Council (EP/R03480X/1, EP/V004859/1); Innovate UK (50414).

\clearpage
\newpage
\appendix 

\setcounter{table}{0}
\renewcommand{\thetable}{A\arabic{table}}%
\setcounter{figure}{0}
\renewcommand{\thefigure}{A\arabic{figure}}%

\section*{Appendices}

\section{In-situ optical measurement system}
\subsection{Optical system}
\begin{figure*}[b]
    \centering
    \includegraphics[width = 0.33\textwidth]{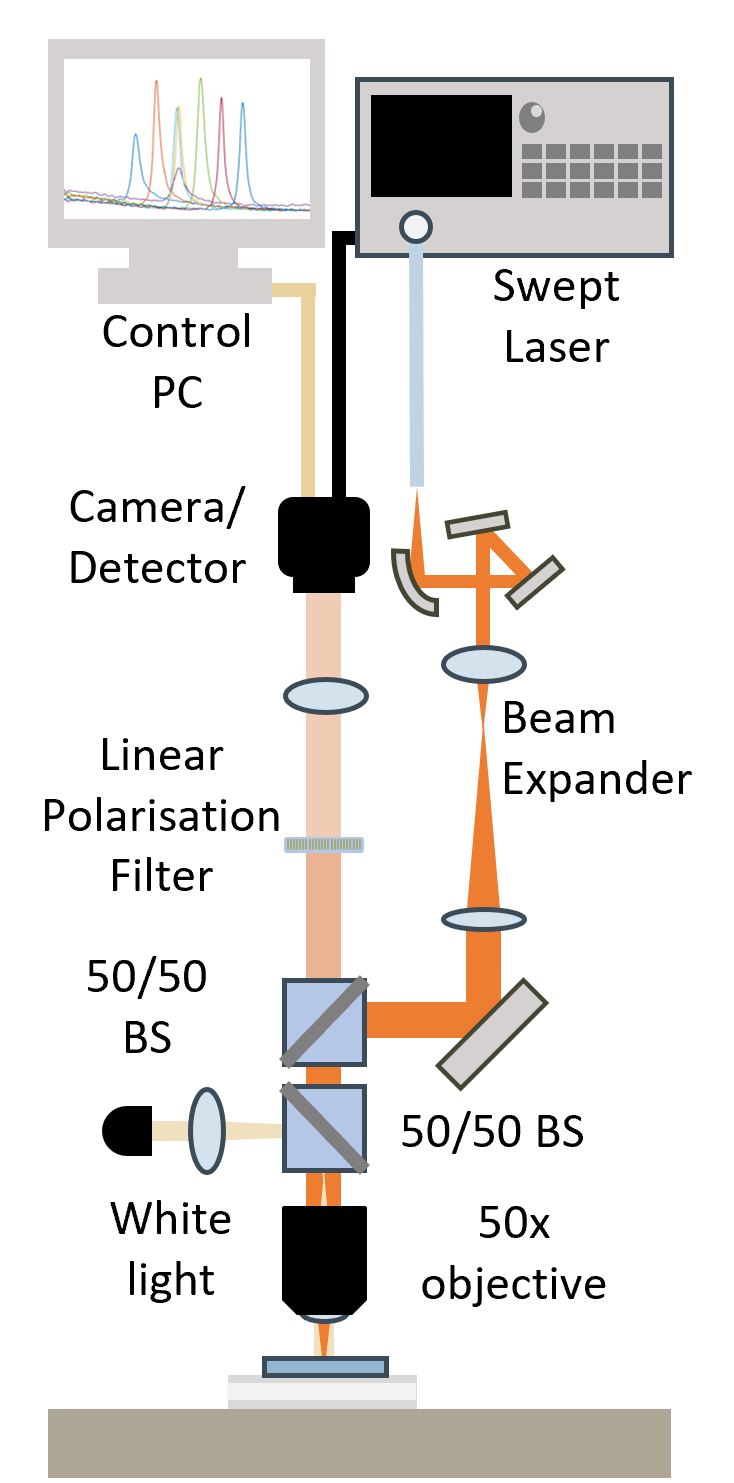}
    \caption{Schematic of the optical measurement system, comprising tuneable laser source and broadband white light illumination, galvo beam steering, microscope column, polarisation filter, and camera detection system.}
    \label{figS:OpticalSchematic}
\end{figure*}

A schematic of the in-situ optical measurement system is shown in \figurename\,\ref{figS:OpticalSchematic}.  
The swept tuneable laser source, Agilent 8614B, is coupled through polarisation maintaining fibres to a parabolic fibre to free space collimating mirror.  The collimator is directly coupled to a scanning galvonometer and lens system for beam steering at the sample plane.  The laser source has a linear polarisation with $>20$\,dB extinction and is aligned at a $45^o$ angle to the x-axis of the translation stage and the mounted samples.  The polarisation filter in the reflected beam path is set at an angle of $90^o$ with respect to the input polarisation to reject specular reflection from the sample surface and microscope column optics.  The polarisation of the PhCC mode is then at $45^o$ with respect to both the input polarisation and output polarisation angle, inducing a 6dB loss in throughput, but suppressing the specular reflection background.  The reflected signal is then focussed on an InGaAs focal plane array for detection. The InGaAs array is housed inside an Lucid Vision Triton Global Shutter camera, the bandgap of the sensor has been optimised to extend the sensitivity down to visible wavelengths, facilitating a broad range of $400 - 1700\,$nm. This allows co-alignment of wide field illuminated device structures to the injected tuneable IR beam using the same sensor.

\subsection{Spectral measurement method}
The spectral measurements of PhCCs were based on swept tuneable laser sampling.  The tuneable laser source was an Agilent 8164A system, running in continuous sweep mode over a $10\,$nm wavelength range around $1550\,$ nm at a constant sweep speed of $0.5\,$nm$/$s. The reflected optical signal from the cavity was measured on the InGaAs focal array, imaging at a maximum frame rate of $833.3\,$fps.  The camera integration time was set by the frame rate and the gain was adjusted to avoid pixel saturation at the maximum reflected signal level. To calculate the reflection intensity in a single camera frame sample period, a region of interest was defined in the image, which remained constant across all frames, and a summation carried out across the pixels in that area.  Representative scans of sample PhCCs in an array were carried out to avoid image saturation, whilst maintaining effective use of the camera's dynamic range.  The signal from each frame was then translated through time sampling to a corresponding laser wavelength.  Given a laser sweep speed of $0.5 nm/s$ and an imaging frame rate of $833.3\,$fps, the measurement spectral resolution was $0.6\,$pm.  This could be improved for higher Q-factor cavity systems by using a slower laser sweep rate.  For dynamic measurements the laser scan range was reduced to $2.5\,$nm to enable faster measurements.


\subsection{Extraction of resonant wavelength and cavity Q-factor}

The resonance wavelength and linewidth were evaluated by fitting the measured reflectivity spectra data around a resonance to an analytical model. Following common practice in literature, we have used a Fano resonance model for fitting our photonic crystal cavity spectra \cite{galli2009light,wu2014spectrally,yang2014all,limonov2017fano}. After removing the low frequency background modulation from the signal, we fit the reflected intensity, $I(\lambda)$, with the Fano equation \cite{limonov2017fano} shown in Eq.[\ref{eq:Fano}]. The fitting was carried out using a nonlinear regression method in MATLAB.

\begin{equation} \label{eq:Fano}
I(\lambda) = A + B\frac{(cot(\delta)+2(\lambda-\lambda_R)/\gamma)^2}{1+(2(\lambda-\lambda_R)/\gamma)^2}
\end{equation}

Here, $\lambda_R$ is the resonant wavelength, $\gamma$ is the resonance width, and $A$ and $B$ are constants relating to the background and peak height respectively. The parameter controlling the deviation from Lorentzian lineshape is $\delta$, which refers to the phase shift from the coupling of a discrete resonant mode to a continuum band of states \cite{limonov2017fano}. It is related to the Fano parameter, $q$, through $q = cot\delta$, and it can be shown that for $q\to0$ or $q\to\pm\infty$ Lorentzian lineshapes are recovered. The Q factors are measured as $Q = \lambda_R/\gamma$.

\begin{figure}
    \centering
    \includegraphics[width=0.7\textwidth]{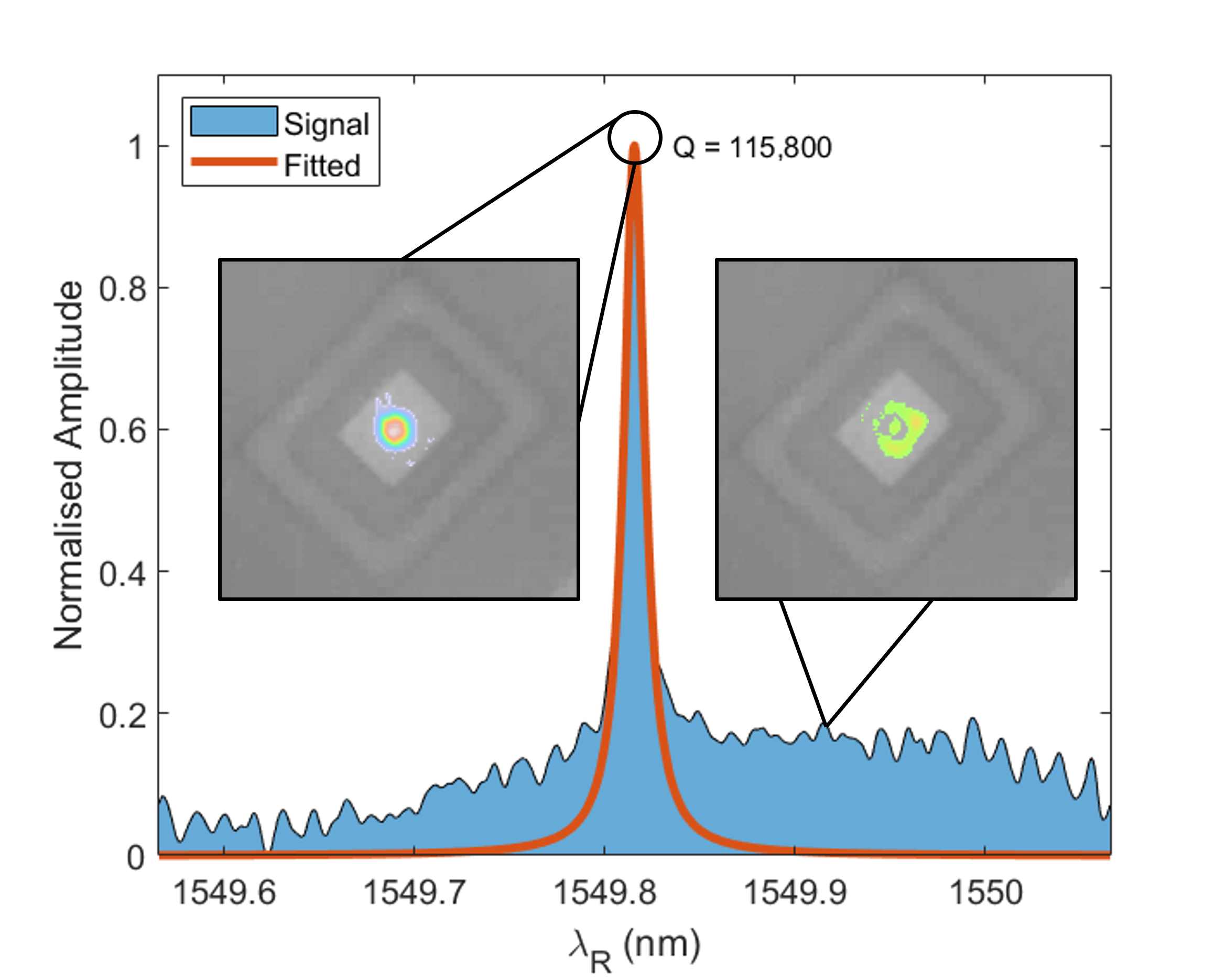}
    \caption{Example of Fano Fitting algorithm for a 2nd Receiver Resonance}
    \label{fig:fano-fitting}
\end{figure}

\subsection{Cavity Q-factors effects}
\begin{figure}
    \centering
    \includegraphics[width=10cm]{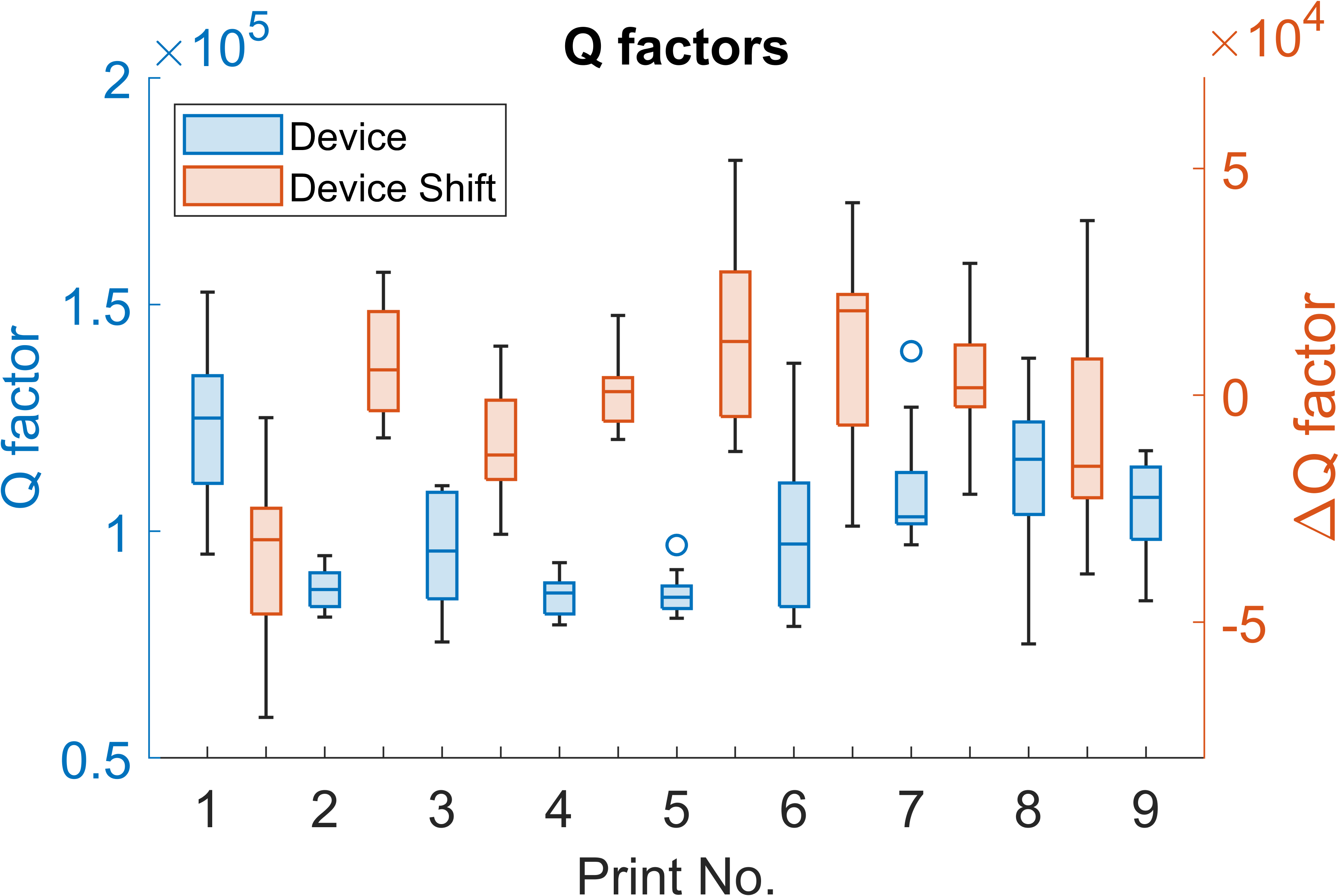}
    \caption{Q-factor variation as a function of repeated print cycle for a set of 10 PhCC devices.  Absolute Q-factor value for the 10 device set is presented along with the device specific Q-factor variation associated with each print cycle.}
    \label{fig:Qfactors}
\end{figure}

In addition to the main results presented in the manuscript that relate to the cavity resonant wavelength, cavity Q-factors are another important element in the transferred device performance.  The Q-factor was calculated as detailed above for the set of 10 cavities that were printed over 9 successive cycles, corresponding to Figure 5 in the main text.  \figurename\,\ref{fig:Qfactors} shows the measured Q-factor of the cavities for each print cycle along with the relative difference in Q-factor for devices between cycles.  The average Q-factor is in the $10^5$ range, and although individual device variations between print cycles can be in the $10^4$ range, the average value is relatively stable over the full set of print cycles.  So although Q-factor is more susceptible to variation than the cavity resonant wavelength, devices remain within a consistent operation range, allowing targeting with resonant wavelength as primary characteristic.

\section{Spatial clusters of cavity resonances within ordered arrays}
\begin{figure*}
    \centering
    \includegraphics[width=10cm]{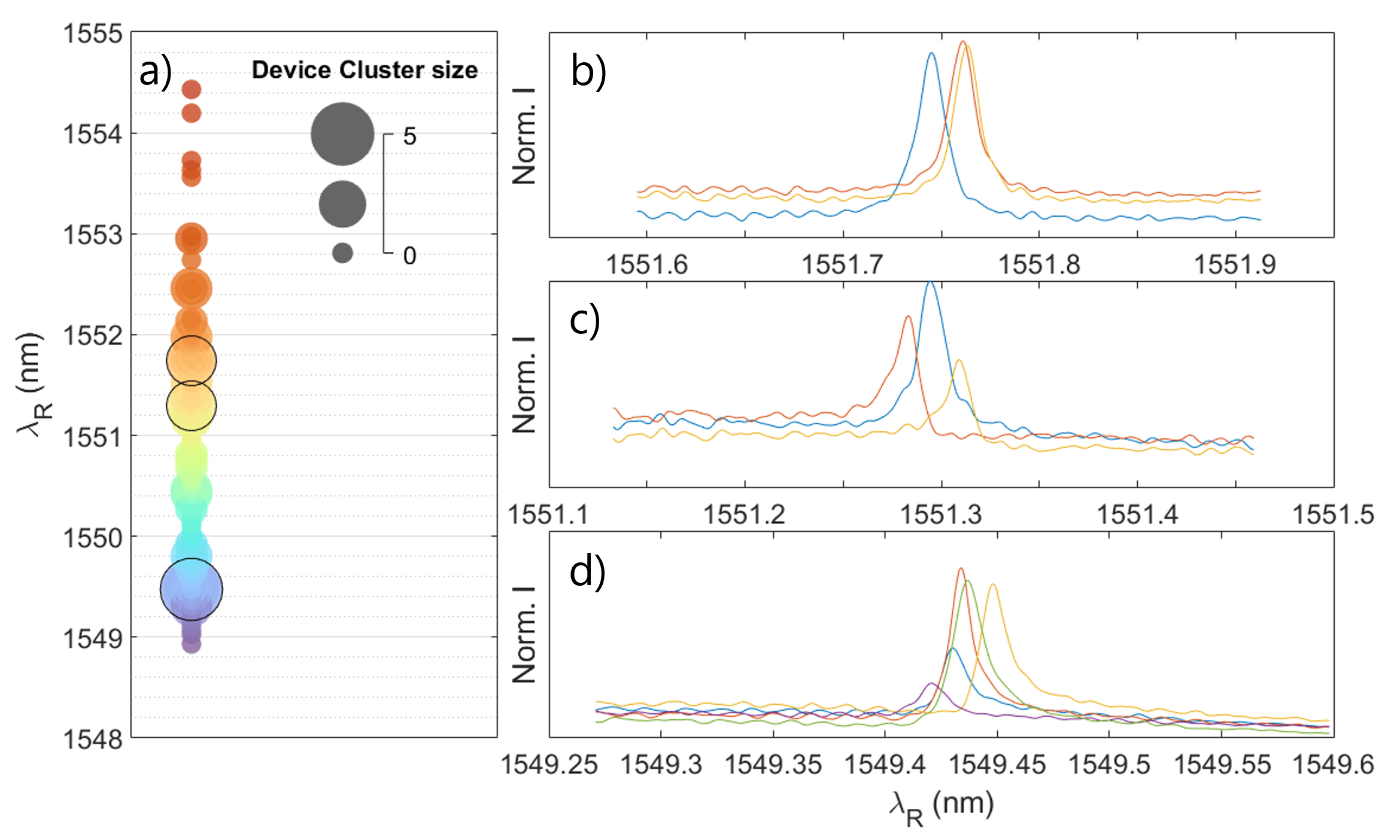}
    \caption{\textbf{Cavity Clustering.} (a)  On the left a bubble chart showing resonant wavelength ordered on the y-axis, spatial position by colour and number of devices within a standard deviation of $0.015\,$nm by bubble size. (b-d) Selected groupings of devices outlined in black in (a) are shown as reflectivity spectra to show the spectral overlap between devices.}
    \label{figS: Cavity Clustering}
\end{figure*}

The ordered arrays of devices maintained the overall standard deviation of resonant wavelength, $\lambda_R$, from the as-fabricated PhCC devices.  The ordering allows for local spatial clusters of devices with closely spaced resonant wavelengths to be assessed.  The average cavity linewidth of $\sim0.015\,$nm was taken as a first approximation of the maximum deviation between individual cavity resonant wavelengths that would allow them to interact with a single narrow linewidth laser source.  \figurename\,\ref{figS: Cavity Clustering} (a) shows the distribution of PhCC clusters across the range of measured resonant wavelengths.  The bubble size is related to the number of devices with resonant wavelengths within the average cavity linewidth from one another.  There are a number of groupings on the order of 5-6 cavities, showing that even without post-fabrication tuning device selection and spatial binning can be achieved even within a set of only 119 devices.  \figurename\ref{figS: Cavity Clustering}(b-d) shows examples  of cavity resonance clusters around a wavelength of $1550 nm$.

\bibliography{PhCC} 

\end{document}